\documentclass[twocolumn,showpacs,preprintnumbers,amsmath,amssymb,floatfix,nofootinbib]{revtex4}

\usepackage{graphicx}
\usepackage{dcolumn}
\usepackage{bm}

\newcommand{\pom}{\tt I\! P}
\newcommand{\beq}{\begin{equation}}
\newcommand{\eeq}{\end{equation}}

\def \pom {{I\!\!P}}

\begin{document}

\title{Exclusive glueball production in high energy nucleus-nucleus collisions}

\author{M. V. T. Machado$^1$ and M. L. L. da Silva$^2$}
\affiliation{$^1$ High Energy Physics Phenomenology Group, GFPAE  IF-UFRGS \\
Caixa Postal 15051, CEP 91501-970, Porto Alegre, RS, Brazil\\
$^2$ Centro de Ci\^encias Fisicas e Matem\'aticas.  Departamento de Fisica - UFSC\\
Bairro Trindade. Caixa Postal 476. CEP 88040-970. Florianopolis. SC. Brazil}

\begin{abstract}
The cross sections for the glueball candidates production in quasi-real photon-photon collisions and on central diffraction processes, i.e. double Pomeron exchange, in heavy ion interactions at RHIC and LHC are computed. The rates for these distinct production channels are compared and they may be a fruitful approach to the investigation of glueballs.
\end{abstract}


\maketitle

\section{Introduction}

The gluon self-coupling in QCD opens the possibility of existing bound states of pure gauge fields known as glueballs. Glueballs ($G$) are predicted by several theoretical formalisms and by lattice calculations. For a comprehensive review on the current status of theoretical and experimental aspects of glueball studies we quote Ref. \cite{Vento} and Ref. \cite{Crede}, respectively.  Many mesons  have stood up as good candidates for the lightest glueball in the spectrum and in particular the scalar sector ($J^{PC} = 0^{++}$) seems promising. The mesons $f_0(1500)$ and the $f_0(1710)$ have been considered  the principal candidates for the
scalar glueball\cite{aleph,closekirk}. However, in
this mass region the glueball state will mix strongly with nearby $q\bar{q}$ states \cite{closekirk,amsler}. More recently, the BES collaboration observed a new resonance called $X(1835)$ \cite{bes}. It is an important candidate for glueball and the nature of meson $X(1835)$ has several interpretations. One of them consider it a pseudo-scalar glueball ($J^{PC} = 0^{-+}$) as first suggested in Ref. \cite{Kochelev} and afterwards in \cite{bing}. 
 
Recently, the clean topologies of exclusive particle production in electromagnetic interactions hadron-hadron
and nucleus-nucleus collisions mediated by colorless exchanges such the QCD Pomeron or two photons have attracted
an increasing interest \cite{upcs}. The cross sections for these processes are smaller than the correspondent
inclusive production channels, which it is compensated by a more favorable signal/background relation.
Experimentally, exclusive events are identified by  large rapidity gaps on both sides of the produced central
system and the survival of both initial state particles scattered at very forward angles with respect to the beam.

Here, we will focus on exclusive glueball production in two-photon and Pomeron-Pomeron interactions in coherent
nucleus-nucleus collisions at high energy colliders (RHIC and LHC). In these cases, the photon flux scales as the
square charge of the beam, $Z^2$, and then the corresponding cross section is highly enhanced by a factor
$\propto Z^4\approx 10^7$ for gold or lead nuclei. A competing channel, which produces similar final state
configuration, is the central diffraction  (CD) process. Such a reaction is modeled in general by two-Pomeron
interaction. Experimentally, the separation of these channels is somewhat difficult and  from theoretical point of
view the Pomeron-Pomeron are subject to large uncertainties at collider energies. One goal of present work is to
compare the cross sections for these two channels in the production of glueball candidates. This paper is organized
as follows: in next  section we present the main expressions for cross section calculation of two-photon and
Pomeron-Pomeron processes and in last section we shown the numerical results and discussions.

\section{Cross section calculation}

Let us start with the glueball production in photon-photon scattering at coherent heavy ion collisions using the
Weizs\"acker - Williams approximation (EPA approximation). In such an approach, the cross section for a two
quasi-real photon process to produce a glueball state, $G$, at center-of-mass energy $W_{\gamma \gamma}$ factorises
into the product of the elementary cross section for $\gamma \gamma \rightarrow G$ convoluted with the equivalent
photon spectra from the colliding ions \cite{upcs}:
\begin{eqnarray}
\sigma_{\gamma\gamma} (AA \rightarrow A+G+A) = \int \frac{dk_1}{k_1} \frac{dk_2}{k_2} \frac{dn_{\gamma}}{dk_1}
\frac{dn_{\gamma}}{dk_2}\sigma\,(\gamma\gamma\rightarrow G), \label{sigfoton}
\end{eqnarray}
where $k_{1,2}$ are the photon energies and $dn/dk$ is the photon flux at the energy $k$ emmited by the hadron $A$.
The photon energies determine the center-of-mass energy $W_{\gamma \gamma}= \sqrt{4k_1k_2}$ and the rapidity $Y$ of
the produced system. Namely, one has $k_{1,2}=(W_{\gamma\gamma}/2)\exp (\pm Y )$ and $Y=(1/2)\ln \,(k_1/k_2)$.  In
addition, $\sqrt{s_{NN}}$ is the center-of-mass energy of the ion-ion system and the Lorentz relativistic factor  is
given by $\gamma_L= \sqrt{s_{NN}}/(2m_N)$. In particular, in the numerical calculations we use
$\sqrt{s_{NN}}=0.2\,(5.5)$ TeV  and $\gamma_L= 109\,(2930)$ for RHIC (LHC).

In the EPA approximation, the flux of equivalent photons from a relativistic particle of charge $Z$ is determined
from the Fourier transform of its electromagnetic field. For an extended charge with electromagnetic form factor,
$F_A(Q^2)$, the energy spectrum can be computed as,
\begin{eqnarray}
\frac{dn_{\gamma/A}\,(x)}{dk} = \frac{\alpha\,Z^2}{\pi}\,\frac{A(x)}{x}\int\frac{Q^2-Q_{\mathrm{min}}^2}{Q^4}
\,|F_A(Q^2)|^2\,dQ^2,
\label{fotflux}
\end{eqnarray}
where $x=k/E$ is the fraction of the beam energy carried by the photon and $A(x)=1-x+(1/2x^2)$. Moreover,
$\alpha=1/137$ and $Q^2$ is the four-momentum transfer squared from the charge, with
$Q_{\mathrm{min}}^2\approx (x m_N)^2/(1-x)$.

The glueball production in two-photon fusion can be calculated using the narrow resonance approximation \cite{BKT}:
\begin{eqnarray}
\sigma\,(\gamma\gamma\rightarrow G) = (2J+1)\,\frac{8\pi^2}{M_G}\,\Gamma(G\rightarrow \gamma \gamma)\,\delta 
\left(W_{\gamma\gamma}^2 -  M_G^2  \right),
\label{twophotres}
\end{eqnarray}
where $\Gamma(G\rightarrow \gamma \gamma)$ is the partial two-photon decay width of $G$, $M_G$ is the glueball mass
and $J$ is the spin of the state $G$. Here, we compute the production rates for the mesons $f_{0}(1500)$,
$f_{0}(1710)$ and $X(1835)$ \cite{PDG}, respectively. The reason is due to they  have been mentioned as possible glueball candidates by phenomenologists \cite{Vento,Crede}.

Some important comments are in order. The predictions for the two-photon component are in practice somewhat difficult as the branching ratios have not been measured. To compute numerical values for the meson (glueball) cross section in two-photon reactions estimates for the two-photon decay widths are needed. The determination of them depend upon whether the meson state is a pure quarkonium, pure gluonic or a mixed hybrid state. For a pure quarkonium state the width can be related (at leading order) to the two-gluon width, $\Gamma(q\bar{q}\rightarrow gg)$. Namely, $\Gamma (q\bar{q}\rightarrow \gamma\gamma ) \simeq  D_c\,e_q^4\,(\alpha/\alpha_s)^2\,\Gamma(q\bar{q}\rightarrow gg)$, where $D_c=9/2$ is the colour factor and $e_q$ is the relevant quark charge. One can estimate the two-gluon width from the total width for the meson state and the theoretical expectation that the $q\bar{q}\rightarrow gg$ branching ratio\footnote{For pure glueball resonance, $G$, the branching ratio is $Br(G\rightarrow gg) \simeq 1$, whereas mixing states  will give intermediate values of branching ratio.} is of order $\alpha_s^2$ \cite{Farrar}. In case of a pure gluonic state, the two-photon width can be computed using a nonrelativistic gluon bound-state model as performed for instance in Ref. \cite{Kada}. There, the unknown parameters as the digluon wavefunction, or its first/second derivative at the origin, are determined by using measured values of $\Gamma (J/\psi \rightarrow G\gamma)$.  

Now, we compute estimates for the two-photon widths assuming pure $q\bar{q}$ and pure gluonic resonances, respectively. For the first case, as discussed above, we take $\Gamma (R\rightarrow \gamma \gamma)= e_q^4\,(3\alpha)^2\,\Gamma_{tot}(R)/2$. Using the Particle Data Group (PDG) average values for the total width one gets $\Gamma (f_0(1500)\rightarrow \gamma\gamma)\simeq 0.3$ keV, $\Gamma (f_0(1710)\rightarrow \gamma\gamma)\simeq 0.4$ keV and $\Gamma (X(1835)\rightarrow \gamma\gamma)\simeq 0.2$ keV. The corresponding cross sections using these theoretical estimates for the width are 3(158) $\mu$b, 3.4(216) $\mu$b and 1.1(84) $\mu$b at RHIC(LHC).  If we are conservative, one can consider the experimental upper bounds for the two-photon widths of $f_0(1500)$ and $f_0(1710)$. This procedure gives an upper limit of the cross section for those resonances in  peripheral collisions. The ALEPH experiment \cite{aleph}  studied the $\gamma\gamma$ production of those glueball candidates via their decay to $\pi^+\pi^-$ and the following limits\footnote{Here, we consider the ALEPH limits $\Gamma (\gamma\gamma\rightarrow f_0(1500))\cdot Br\,(f_0(1500)\rightarrow \pi^+\pi^-)<0.31$ keV, $\Gamma (\gamma\gamma\rightarrow f_0(1710))\cdot Br\,(f_0(1710)\rightarrow \pi^+\pi^-)<0.55$ keV and taking the branching ratios $0.30\pm 0.07$ and $0.026\pm 0.016$ \cite{aleph}, respectively } were determined: $\Gamma(f_0(1500) \to \gamma \gamma) \leq 1.08$ keV and $\Gamma(f_0(1710) \to \gamma \gamma) \leq 21.25$ keV. Using those limits the corresponding cross sections are of order 0.95 mb (20 $\mu$b) for $f_0(1500)$ and 11.5 mb (180 $\mu$b) for $f_0(1710)$ at LHC (RHIC) energies. We quote Ref. \cite{BertulaniGamma} for a comparison of our results with a wide class of theoretical models and exotic QCD states in the meson production in photon-photon process. 

For a pure glueball resonance we follow \cite{Kada}, adapted for the candidates considered here. Namely, assuming the $f_0$ resonances as states with $J=L=0$ then  Eq. (54) of Ref. \cite{Kada} has been used, where we take the PDG values for the radiative $J/\psi$ decays in the following channels:  $\psi\rightarrow \gamma f_0(1500)\rightarrow \gamma \pi\pi $ and $\psi\rightarrow \gamma f_0(1710)\rightarrow \gamma K\bar{K}$. Assuming the $X(1835)$ resonance to be a state with $J=0$ and $L=S=1$ we relay on Eq. (35) of \cite{Kada} and use the PDG value for the decay channel $\psi\rightarrow \gamma X(1835)\rightarrow \gamma \pi^+\pi^-\eta^{\prime}$.  Putting all together, the estimates for the two-photon width for a pure glueball resonance are $\Gamma(f_0(1500) \to \gamma \gamma) \simeq 0.77$ eV, $\Gamma(f_0(1710) \to \gamma \gamma) \simeq 7.03$ eV  and $\Gamma(X(1835) \to \gamma \gamma) \simeq 0.021$ keV. Notice that for the $X(1835)$ a larger width is predicted \cite{bing}, being of order 1.1 keV. The widths are about three orders of magnitude smaller that for pure $q\bar{q}$ states.  Therefore, as the two-photon cross section scales as $(2J+1)\Gamma \,(R\rightarrow \gamma\gamma)$, Eq. (\ref{twophotres}), one can consider the experimental feasibility of using peripheral heavy-ion collisions to determine the nature of the resonances discussed above.  The values for the corresponding widths and corresponding cross sections estimates are shown in Table I.

Now, we address the Pomeron-Pomeron channel. In particular, we focus on the central diffraction (double Pomeron
exchange, DPE) in nucleus-nucleus interactions. As a starting point we compute the DPE proton-proton cross section
making use of the  Bialas-Landshoff  \cite{Land-Nacht,Bial-Land} approach. We believe that this non-perturbative
approach is a reasonable choice due to the light mass of glueballs candidates considered in present calculation.
For a perturbative QCD guided calculation we quote the recent work in Ref. \cite{Antony}, where the exclusive
scalar $f_0(1500)$ meson production is carefully investigated.  Here, we are particularly interested in the exclusive
and central inclusive (central inelastic) DPE production of glueball states. In the exclusive DPE event the central
object $G$ is produced alone, separated from the outgoing hadrons by rapidity gaps,
$pp\rightarrow p+\text{gap}+G+\text{gap}+p$. In the central inclusive DPE event an additional radiation accompanying
the central object is allowed. In approach we are going to use, Pomeron exchange corresponds to the exchange of a
pair of non-perturbative gluons which takes place between a pair of colliding quarks. For DPE central inclusive $G$
production we can neglect the additional gap spoiling effect, so-called Sudakov effect. The scattering matrix is given by,
\begin{eqnarray}
\mathcal{M} & = & \mathcal{M}_{0}\left(  \frac{s}{s_{1}}\right)  ^{\alpha(t_{2})-1}\left(  \frac{s}{s_{2}}\right)
^{\alpha(t_{1})-1}\,F(t_{1})\,F(t_{2})\nonumber \\
&\times & \exp\left(  \beta\left(  t_{1}+t_{2}\right)  \right)\,  S_{\text{gap}}\left(\sqrt{s} \right).
\label{M_all}
\end{eqnarray}
Here $\mathcal{M}_{0}$ is the amplitude in the forward scattering limit ($t_1=t_2=0$). The standard Pomeron Regge
trajectory is given by $\alpha\left(  t\right)=1+\epsilon+\alpha^{\prime}t$ with
$\epsilon\approx 0.08,$ $\alpha^{\prime}=0.25$ GeV$^{-2}$. The momenta of incoming (outgoing) protons are labeled by $p_1$
and $p_2$ ($k_1$ and $k_2$), whereas the glueball momentum is denoted by $P$. Thus, we can define the following quantities
appearing in Eq. (\ref{M_all}): $s=(p_{1}+p_{2})^{2}$, $s_{1}=(k_{1}+P)^{2},$ $s_{2}=(k_{2}+P)^{2},$
$t_{1}=(p_{1}-k_{1})^{2}$ , $t_{2}=(p_{2}-k_{2})^{2}$. The nucleon form-factor is given by $F_p\left(  t\right)  $ = $\exp(b
t)$ with $b=$ $2$ GeV$^{-2}$. The phenomenological factor $\exp\left(  \beta\left(  t_{1}+t_{2}\right)  \right)
$ with $\beta$ $=$ $1$ GeV$^{-2}$ takes into account the effect of the momentum transfer dependence of the non-perturbative
gluon propagator. The factor $S_{\text{gap}}$ takes the gap survival effect into account $i.e.$ the probability
($S_{\text{gap}}^{2}$) of the gaps not to be populated by secondaries produced in the soft rescattering.  For our purpose
here, we will consider $S_{\mathrm{gap}}^2=0.032$ at $\sqrt{s}=5.5$ TeV in nucleon-nucleon collisions \footnote{It is obtained using  a parametric interpolation formula for the KMR survival probability factor \cite{KKMR} in the form
$ S_{\mathrm{gap}}^2  =a/[b+\ln (\sqrt{s/s_0})]$ with $a = 0.126$, $b=-4.988$ and $s_0=1$ GeV$^2$. This formula interpolates
between CD survival probabilities of $4.5 \, \%$ at Tevatron and $2.6 \, \%$ at the LHC.} and $S_{\mathrm{gap}}^2=0.15$ at
$\sqrt{s}=200$ GeV (RHIC). In particular, for RHIC we have used an estimation using a simple one-channel eikonal model for the survival probability \cite{GLM}, whereas for the LHC energy we follows Ref. \cite{KKMR} that considers a two-channel eikonal model that embodies pion-loop insertions in the pomeron trajectory, diffractive dissociation and rescattering effects. We quote Ref. \cite{GLMrev} for a detailed comparison between the two approaches and further discussions on model dependence of inputs and consideration of multi-channel calculations.

Following the calculation presented in Ref. \cite{Bial-Land} we find
$\mathcal{M}_{0}$ for colliding hadrons,
\begin{eqnarray}
\mathcal{M}_{0}=32 \,\alpha_0^2\,D_{0}^{3}\,\int d^{2}\vec{\kappa}\,p_{1}^{\lambda}V_{\lambda\nu}^{J}p_{2}^{\nu}\,
\exp(-3\,\vec{\kappa }^{2}/\tau^{2}),
\label{M_o}
\end{eqnarray}
where $\kappa $ is the transverse momentum carried by each of the three gluons. $V_{\lambda\nu}^{J}$ is the 
$gg\rightarrow G^{J}$ vertex depending on the polarization $J$ of the $G^{J}$ glueball meson state. For the
cases considered here, $J=0$, one obtains the following result
\cite{Bial-Land,KMRS}:
\begin{equation}
p_{1}^{\lambda}V_{\lambda\nu}^{0}p_{2}^{\nu}=\frac{s\,\vec{\kappa}^{2}
}{2M_{G^{0}}^{2}}A, \label{p_1Vp_2}%
\end{equation}
where $A$ is expressed by the mass $M_{G}$ and the width $\Gamma (gg\rightarrow G)$ of the glueball meson through the relation:
\begin{equation}
A^{2}= 8\pi M_G\,\Gamma (gg\rightarrow G). \label{A^2}
\end{equation}
For obtaining the two-gluon decays widths the following relation is used,
$\Gamma \,(G\rightarrow gg)=\mathrm{Br}\,(G\rightarrow gg)\,\Gamma_{tot}(G)$. At this point, some discussion is in order. The two-gluon width depends on the branching fraction of the resonance $R$ to gluons, $\mathrm{Br}\,(G\rightarrow gg)$ and its knowledge would give quantitative information on the glueball content of a particular resonance. As discussed before, it is a theoretical expectation \cite{Farrar} that $\mathrm{Br}\,(R(q\bar{q})\rightarrow gg)={\cal O}(\alpha_s^2)\simeq 0.1-0.2$ whereas $\mathrm{Br}\,(R(G)\rightarrow gg)\simeq 1$. Here, we will be conservative and assume the resonance to be a pure glueball. This fact translates into an upper bound for the exclusive DPE production as the cross section scales with $\Gamma\,(R\rightarrow gg)$. Following Ref. \cite{close}, the two-gluon width can be computed from the resonance branching fraction in $J/\psi$ radiative decay, $\mathrm{Br}\,(\psi\rightarrow \gamma \,G)$. For the candidates of interest here one obtains:
\begin{eqnarray}
\mathrm{Br}\,(G(0^{++})\rightarrow gg) = \frac{8\pi(\pi^2-9)\,Br[\psi\rightarrow \gamma\,G(0^{++})]}{c_R\,x|H_J(x)|^2\,\Gamma_{tot}}\frac{M_{\psi}^2}{M_G},\nonumber\\
\mathrm{Br}\,(G(0^{-+})\rightarrow gg) = \frac{8\pi(\pi^2-9)Br[\psi\rightarrow \gamma\,G(0^{-+})]}{c_R\,x|H_J(x)|^2\,\Gamma_{tot}}\frac{M_{\psi}^2}{M_G},\nonumber
\end{eqnarray}
where the function $H_J((x)$ is determined in the non-relativistic quark model (NRQM) (see appendix of Ref. \cite{close}) and $c_R$ is a numerical constant ($C_R=1, \,2/3,\,5/2$ for $J^{PC}=0^{-+},\,0^{++},\,2^{++}$, respectively). The masses of $J/\psi$ and of resonance are $M_{\psi}$ and $M_G$, respectively,  and $x=1-(M_G^2/M_{\psi}^2)$. Based on equations above, in Ref. \cite{close} the following values for the branching fractions for scalar glueballs candidates are obtained:  $\mathrm{Br}\,[f_0(1500)]=0.64\pm 0.11$, $\mathrm{Br}\,[f_0(1710)]=0.52\pm 0.07$. For the pseudoscalar $X$ the situation is less clear due to small information on its decaying channels in radiative $J/\psi$ decays. The authors in  \cite{close} have a prediction for $\eta$ resonance which gives $\mathrm{Br}\,[\eta(1410)]=0.9\pm 0.2$. As the branching fraction scales as $1/M_G$ in this theoretical model, an educated guess for the $X$ branching fraction would be $\mathrm{Br}\,[X(1835)]=(M_{\eta}/M_{X})\cdot \mathrm{Br}\,[\eta(1410)]=0.69 \pm 0.15$. In the numerical calculations  we set the limit case $\mathrm{Br}\,[X(1835)]=1$ and notice that the branching would be about 30 \% smaller. The values for $\Gamma_{gg}$ used in our calculations are summarized in Table II. A consequence on the small deviation for the branching fraction in pure $q\bar{q}$ and glueball resonance is the difficulty in testing their nature using the exclusive diffractive data. An option would be to obtain for instance the differential cross section on angular distributions and then compare the predictions for each composition (pure $q\bar{q}$, mixing state and pure glueball). 

\begin{table}[t]
\centering
\renewcommand{\arraystretch}{1.5}
\begin{tabular}{c c c c}
\hline
 Glueball Candidate & $\Gamma_{\gamma \gamma}$ [eV] & RHIC [nb] & LHC  [$\mu$b] \\
\hline
  $f_0\,(\mathrm{1500})$  & 0.77 &  14-9.3 & 0.7-1.3  \\
  $f_0\,(\mathrm{1710})$  & 7.03 & 60-43 & 3.8-8.6 \\
  $X\,(\mathrm{1835})$  & 0.021 & 0.11-0.09 & 0.01-0.02 \\
\hline
\end{tabular}
\caption{Cross sections for pure glueball candidates production through photon-photon fusion in   electromagnetic nucleus-nucleus collisions at RHIC and LHC energies.}
\label{tab1}
\end{table}

In addition,  we use the parameters $\tau=1$ GeV and $D_{0}G^{2}\tau=30$ GeV$^{-1}$ \cite{Bial-Land} where $G$ is the scale
of the process independent non-perturbative quark gluon coupling.  An indirect determination of unknown parameter
$\alpha_0=G^2/4\pi$ has been found in Ref. \cite{Adam} using experimental data for central inclusive dijet production cross
section at Tevatron. Namely,  it has been found the constraint
$S_{\mathrm{gap}}^2\,(\sqrt{s}=2\,\mathrm{TeV})/\alpha_0^2 = 0.6$, where $S_{\mathrm{gap}}^2$ is the gap survival probability
factor (absorption factor). Considering the KMR \cite{KKMR} value $S_{\mathrm{gap}}^2=0.045$ for CD processes at Tevatron
energy, one obtains $\alpha_0=0.274$.

The calculation presented above concerns to central inclusive process, where the QCD radiation accompanying the produced
object is allowed. Therefore, in order  to describe the exclusive processes where the central object is produced alone
we include the Sudakov survival factor $T(\kappa,\mu)$ \cite{Khoze} inside the loop integral over $\vec{\kappa}$. The Sudakov
factor $T(\kappa,\mu)$ is the survival probability that a gluon with transverse momentum $\kappa $ remains untouched in the
evolution up to the hard scale $\mu=M_G/2$. The function $T(\kappa,\mu)$ is given by \cite{Khoze}:
\begin{eqnarray}
T(\kappa ,\mu) & = & \exp\left(  -\int_{\vec{\kappa}^{2}}^{\mu^{2}
}\frac{\alpha_{s}\left(  \vec{k}^{2}\right)  }{2\pi}\frac{d\vec
{k}^{2}}{\vec{k}^{2}} \right. \nonumber \\
& \times & \left. \int_{0}^{1-\delta}\left[
zP_{gg}\left(  z\right)  +\sum\limits_{q}P_{qg}(z)\right]  dz\right)  ,
\label{T-def}
\end{eqnarray}
where $\delta=\left|  \vec{k}\right|  /(  \mu+\left|  \vec{k}\right| )  $, $P_{gg}\left(  z\right)  $ and $P_{qg}(z)$ are the
DGLAP spitting functions. In next section we will discuss the effect of  introducing the Sudakov factor in the estimation of exclusive production in the Pomeron-Pomeron channel.

In order to calculate the $AA$ cross section the procedure presented in Ref. \cite{Pajares} is considered, where the central
diffraction and single diffraction cross sections in nucleus-nucleus collisions are computed using the so-called
{\it criterion C} (we quote Ref. \cite{Pajares} for further details).  Using the profile function for two colliding nuclei, $T_{AB}=\int d^2\bar{b}\,T_A(\bar{b})\,T_B(b-\bar{b})$, the final expression for CD  cross section in $AA$
collisions is given by \cite{Pajares}:
\begin{eqnarray}
\sigma^{\mathrm{CD}}_{AA} =  A^2\int d^2b \,T_{AA}(b)\,\exp\left[-A^2\,\sigma^{in}_{pp}\,T_{AA}(b)\right] \,\sigma^{\mathrm{CD}}_{pp}.
\label{sdxs1}
\end{eqnarray}
where $\sigma^{in}_{pp}$ and $\sigma^{\mathrm{CD}}_{pp}=S_{\mathrm{gap}}^2\times\sigma^{\mathrm{CD}}_{pp}(\sqrt{s}) $ are the inelastic and CD cross sections in proton-proton case, respectively. 

\begin{table}[t]
\centering
\renewcommand{\arraystretch}{1.5}
\begin{tabular}{l c c c}
\hline
 Glueball & $\Gamma_{gg}$ [MeV] & RHIC [mb] & LHC [mb]\\
\hline
  $f_0(\mathrm{1500})$  & 69.8 & $0.63\pm 0.21$ (inc.) & $0.77\pm 0.51$ (inc.)\\
                        & & $0.40\pm 0.14$            (exc.)  &  $0.50\pm 0.32$   (exc.)          \\
  $f_0(\mathrm{1710})$  & 70.2  & $0.68\pm 0.26$ (inc.) & $0.80\pm 0.52 $ (inc.)\\
                        & & $0.41\pm 0.16$      (exc.)   &  $0.49\pm 0.31$     (exc.)       \\
  $X(\mathrm{1835})$  & 70.27 & $0.64\pm 0.24$ (inc.) & $0.77\pm 0.50$ (inc.) \\
                      & & $0.38\pm 0.14$ (exc.) &  $0.45 \pm 0.29$ (exc.)  \\
\hline
\end{tabular}
\caption{Cross sections for inclusive (inc.) and exclusive (exc.) glueball production in double Pomeron exchange process
for RHIC and LHC energies.}
\label{tab2}
\end{table}

Using Woods-Saxon nuclear densities and considering the inelastic cross section $\sigma^{in}_{pp}=73\,(49) $ mb for LHC (RHIC) energy,
$\sqrt{s_{AA}}=5.5\,(0.2)$ TeV, we compute the CD cross section for nuclear collisions. The values for the inelastic cross section are obtained from $\mathrm{DPMJET}$ \cite{DPM}, where the scattering amplitude is parameterized using $\sigma_{tot}$, $\rho$ and elastic slope (these parameters are taken as fitted by the $\mathrm{PHOJET}$ model \cite{Engel}). We notice that for LHC energy the effective
atomic number dependence is proportional to $A^{1/3}$, which means that the nuclear CD cross section is only one order of magnitude
larger than the nucleon-nucleon cross section. For completeness, we give the values of the DPE cross sections for the proton-proton case used in Eq. (\ref{sdxs1}): $\sigma^{\mathrm{CD}}_{pp}(\mathrm{RHIC})= 0.170,0.180,0.168$ mb and $\sigma^{\mathrm{CD}}_{pp}(\mathrm{LHC})= 0.80,0.85,0.83$ mb for $f_0(1500)$, $f_0(1710)$ and $X(1835)$, respectively.

In next section we compare the two production channels and investigate the main theoretical uncertainties. We provide estimates of
cross sections and event rates for both processes for RHIC and LHC energies at the heavy ion mode.

\section{Results and discussions}

In what follows the numerical results for the two-photon and Pomeron-Pomeron processes are presented and discussed. In Table I the
cross sections for glueball production in photon-photon fusion at RHIC and LHC energies are shown. For RHIC we have considered the nominal center of mass energy of 200 GeV for gold-gold collisions and for LHC we take the planned nominal energy of 5500 GeV in lead-lead collisions. The first value corresponds to the cross
section obtained using a non-factorizable photon flux (Cahn-Jackson)  \cite{KJ} and the second one refers to the factorizable flux as shown in Eqs. (\ref{sigfoton}-\ref{fotflux}). The deviation is sizable for RHIC and LHC.  The cross sections are sufficiently large for experimental measurement. The event rates can be
obtained using the beam luminosity \cite{upcs}: for LHC one has ${\cal L}_{\mathrm{PbPb}} = 5 \cdot 10^{26}$ cm$^{-2}$s$^{-1}$, which
produces the following number of events. One has  $3.6\cdot 10^2$, $2\cdot 10^3$ and $4$ for $f_0(1500)$, $f_0(1710)$ and
$X(1835)$, respectively,  in the nominal LHC running time with ions of $10^6$ s (one month). The event rates can be enhanced in a pPb
mode, where the nominal beam luminosity is increased three order of magnitude compared to the PbPb mode. The present calculation can be
compared to previous studies on glueball production in heavy ion collisions \cite{Natale,Schramm}. In general, the numerical results are
similar to those computations and the main deviation comes from the distinct estimates for the two-photon decays widths. A direct
comparison can be done for the $f_0(1710)$ case, where in Ref. \cite{Schramm} one gets 48 nb for RHIC and 2.3 $\mu$b for LHC
(using cut on impact parameter $b>2R_A$ and using $\Gamma_{\gamma \gamma}\simeq 4$ eV \cite{Schramm}). 

For the convenience of phenomenologists we provide here a parameterization of the ultraperipheral $AA$ cross section as a function of the resonance mass at the LHC energy. This makes simple the computation of event rates provided the specific meson  state and its two-photon decay width. Using the Cahn-Jackson photon flux, we obtain in the interval $400\leq M_R\leq 4000$ MeV the parametrization:
\begin{eqnarray}
\frac{\sigma_{\mathrm{upc}}\, (AA\rightarrow R_J+AA)}{(2J+1)\,\Gamma(R_J\rightarrow \gamma\gamma)}= \frac{\sigma_0\,M^{\beta}_R}{1+\left(M_R/4\right)},
\end{eqnarray}
where $\sigma_0=4.9147$ mb/GeV and  $\beta=-3.45335$; $\Gamma_{\gamma\gamma}$ and $M_R$ are the decay width and  the resonance mass in units of GeV, respectively. Several authors have argued for a low lying scalar glueball, with mass between 500 and 1200 MeV \cite{Vento,Crede}, depending on the proponents. The parameterization above allows to obtain estimates starting from a modeling for the two-photon width.

In Table II the results for Pomeron-Pomeron production of glueball is presented. The estimates are shown for the inclusive (inc.) and
exclusive (exc.) double Pomeron exchange as discussed in previous section. Namely, for the inclusive production the Sudakov survival factor is not included (glueball is produced in association with Pomeron remnants) whereas for the exclusive case it is taken into account. In order to estimate the model dependence in the CD cross
section, we have changed the soft Pomeron parameters in order to be consistent with the semi-hard Pomeron values considered in the
DESY-HERA fits to diffractive deep inelastic scattering (DDIS). For instance, taking FIT A of the H1 Coll. \cite{H1diff} parameterization
for the diffractive structure function $F_2^{D(3)}$ one has $\epsilon = 0.118$, $\alpha^{\prime}=0.06$ and $b=2.75$  GeV$^{-2}$. Such a
change enhances the cross section by a factor 3 for PbPb collisions at the LHC. In Table II, the cross sections are presented taking into
account such a theoretical error band. Lower bound corresponds to soft Pomeron parameters and upper bound stands for the semihard Pomeron
ones. For RHIC energy, the Pomeron-Pomeron contribution seems to be bigger than the photon-photon channel in a large extent. On the other
hand, at the LHC they are competitive. However, the Pomeron contribution can be easily separated from photon channel by imposing a cut on
the impact parameter of collision. After imposing this kinematic cut ($b>2R_A$) the Pomeron contribution is reduced as they are dominated
by small impact parameter contributions. 

The present result is difficult to be compared directly to previous studies on
Refs. \cite{Natale,Schramm}. Those author did not include survival probability gap on their calculations and the theoretical approaches
for Pomeron-Pomeron interaction are distinct. For instance, in Ref. \cite{Natale} the $\pom \pom \rightarrow G$ cross section is obtained
using the Pomeron-quark coupling like a isoscalar photon, which allows to obtain the DPE cross section from the two-photon one. On the other hand, in Ref. \cite{Schramm}, only the inclusive double Pomeron production is considered. Following that study, we can
perform a closer comparison. The cross sections are computed there with inelastic scattering effects using the Glauber approximation
(in Table 3 of Ref. \cite{Schramm}, see $\sigma_{AA}$ elastic), which is similar to procedure presented here. After including gap survival
probability factor one gets for the $f_0(1710)$ meson the values 1.23 (3.04) mb  for RHIC (LHC), which is not so far from our results
presented for inclusive production in Table II.

Finally, it is important to discuss the uncertainties on the current calculations and the experimental feasibility of detecting glueballs candidates. The main uncertainty here is the model dependence on obtaining the two-photon and the two-gluon widths for a pure glueball meson. For the two-photon width we considered a nonrelativistic gluon bound-state model of Ref. \cite{Kada}, which it could be a debatable issue and it is far from being optimal. There are more modern approaches as reviewed in Ref. \cite{Crede}, but this is out of the scope of present work. For the two-gluon widths, we obtained them from the quarkonium width based on a non relativistic boundstate calculation \cite{close}. This type of matrix elements have been discussed in Refs. \cite{Chanowitz} giving rise to an effect of chiral suppression. We did not discuss the implication of those findings in present calculation. Concerning the experimental detection, the advantage of the exclusive processes discussed here is clear: glueballs are probably being produced with a high cross section in inelastic collisions (in $pp$ or $AA$ reactions) but when the multiplicity is high  the combinatorial background is overwhelming. In {\it exclusive} production there is no combinatorial background. In the ultraperipheral two-photon production of glueballs, the final state configuration is clear:  nuclei remain intact after collision and a double large rapidity gap between them is present (glueball is centrally produced with a low $p_T$ transverse momenta spectrum). This type of measurement is already done at RHIC for photoproduction of vector mesons and exclusive dilepton production with a  signal identification well understood \cite{Nystrand}. The situation for DPE glueball production is similar, with the $p_T$ spectrum being broader than the processes initiated by two-photons. Thus, a transverse momentum cut (and also impact parameter of collision) could separate the two channels (for a review on these issues we quote Ref. \cite{Albrow}).

\begin{acknowledgments}
The authors thank Curtis A. Meyer, Nikolai Kochelev,  Pedro Bicudo and Dimiter Hadjimichef for comments/suggestions. One of us (MVTM) acknowledges the Aristotle University of Thessaloniki and the organizers of the {\it Low-$x$ Meeting} (Kavala, Greece.
June 21-27 2010) for their invitation, where this work was accomplished. This research was supported by CNPq, Brazil.
\end{acknowledgments}

\end{document}